\documentstyle[12pt]{article}
\begin{document}
\setlength{\baselineskip}{24pt}
\title{ QCD Evolution by Finite Element Methods}
\author{C.J. Benesh,\\
Bonner Nuclear Laboratory,\\
Rice University,\\
Box 1892, Houston, Texas, 77251\\
\\
and\\
\\
Theoretical Division, MSB283,\\
Los Alamos National Laboratory, \\
Los Alamos, NM 87545}

\maketitle
\vfill\eject

\begin{abstract}
	A simple, new method for solving for the $Q^2$
evolution of
parton distributions in perturbative QCD using cubic
splines
is described and applied to the evolution of nonsinglet
quark distributions.

\end{abstract}
\section{Introduction}
	The efficacy of using perturbative QCD to
describe processes
at large momentum transfer in terms of process
independent parton
distributions has spawned an industry devoted to the
determination
of these distributions\cite{1}. Central to this
endeavour is the ability
to solve the Altarelli-Parisi equations\cite{7}, which
govern the $Q^2$
evolution of parton distributions, so that data taken at
different energies
can be used to constrain the form of the parton
distributions. The evolution
equations have been solved in a variety of ways, ranging
from brute force
integration of the integro-differential
equations\cite{4}, to orthogonal
polynomial expansions\cite{5}, to
solutions in terms of Bernstine polynomials for next to
leading order
evolution\cite{6}. In this report,
a straightforward new method for solving the
Altarelli-Parisi
equations
in terms of a small number of cubic splines is described
and illustrated for
the case of nonsinglet quark distributions. Since the
method relies
only on the smoothness of the parton distributions and
the moment structure
of the QCD evolution equations, it is applicable to
singlet and nonsinglet
distributions in both leading and next to leading order
with equal ease.

\section{The Method}

	In this section, the application of finite
element methods to the
problem of QCD evolution is demonstrated for the simple
case of a nonsinglet
quark distribution.
Given a nonsinglet quark distribution $q_{NS}(x,Q_0^2)$
at some initial
renormalization scale $Q_0^2$, we are looking for a new
distribution
evolved to a different $Q^2$ according to QCD
perturbation theory.
To leading order, this requires
\begin{equation}
{\cal M}_n(Q^2) = {\cal M}_n(Q_0^2)\times
\Big{(}{\log(Q_0^2)\over \log(Q^2)}
\Big{)}^{d^{NS}_n},
\end{equation}
where ${\cal M}_n(Q^2)=\int_0^1\,dx\, x^{n-1}$\,
$q_{NS}(x,Q^2)$
is the nth moment of
the quark distribution, and $d^{NS}_n$ is the anomalous
dimension of the
nth nonsinglet, twist two operator.

In principle, knowledge of all the moments is required
to uniquely determine
$q_{NS}(x,Q^2)$, but the distributions used in QCD
phenomenology
are peaked at small values of $x$, and all but a few of
the moments are
small. Exploiting this fact, we proceed by numerically
calculating
the first N moments of the desired distribution in terms
of a set of
N suitably chosen integration points and weights. Thus,
\begin{equation}
{\cal M}_n(Q^2)\approx\sum_{i=1}^N x_i^{n-1}w_i
q_{NS}(x_i,Q^2),
\end{equation}
where $\{x_i\}$ are the integration points and $\{w_i\}$
are the
weights. Assuming that the integrals are well
approximated, this yields
a matrix equation relating the first $N$ moments of the
distribution
to the values of the distribution function at the points
$\{x_i\}$.
In principle, the distribution can be solved for by
taking the
inverse of the matrix and the limit as the number of
integration
points gets large. As a practical method, this procedure
fails
since the matrix,
\begin{equation}
A_{ni}=x_i^{n-1}w_i,
\end{equation}
is difficult to invert numerically when N is large.
For relatively small values of $N$, however, $A_{ni}$
can be inverted easily
and quickly by any one of a variety of standard
techniques. The result
is an approximation of the values of the parton
distribution at $N$ points.
All that is left is to fill in the rest of the curve
between these points.
Since the distribution is expected to be smooth and
non-oscillatory,
the remainder of the curve is found by interpolating
between the
points $x_i$ using cubic splines\cite{8}.
Since the functions of interest are
singular at $x=0$, some care must be taken when choosing
the integration
scheme and spline basis so that the function may be
adequately fit at small
$x$. While other techniques for dealing with this
problem exist in the
literature\cite{8}, here we simply change the variable
in the
numerical integration
and spline fit to $y=x^\alpha$, with $\alpha < 1$, so
that
small values of $x$ have increased weight in the
integrations and the curve
is better approximated by a polynomial  in $y$ near
$x=0$.

\section{Results and Discussion}

	The essence of the method outlined in the
previous section is
the reconstruction of an essentially arbitrary curve
from its moments.
The simplest test available is to decompose a curve into
moments and
then reconstruct it without any change in $Q^2$. In
Figure 1, the results are
displayed for a curve typical of those encountered in
QCD phenomenology,
$xq_{NS}(x)=\sqrt{x}(1-x)^3$, using a Gauss-Legendre
integration scheme
with $\alpha =2/3$, N=3,5,7 or 10, and splines that are
functions of
$y=\sqrt{x}$. For N=3 the procedure has clearly failed
to reproduce the
original, while for N=5 the reconstructed curve deviates
only slightly.
For N=7 and 10, however, the reconstructed curves are
virtually
indistinguishable from the original. In Figure 2, a set
of 10 splines is used
to perform the leading
order evolution of the valence d quark distribution
parametrized by Morfin
and Tung(MT)\cite{4} from an initial scale $Q_0^2=4$
GeV$^2$ to $Q^2=$ 10 and 50
GeV$^2$. Using $\alpha= 1/2$ for both the integration
points and splines,
the MT curves are well reproduced. (The small deviation
from the MT curve
is a reflection of the approximate nature of the
parametrization of the
$Q^2$ dependence of $xd_V(x)$ in reference 4.)

	The finite element method for solving the
Altarelli-Parisi equations
has a number of conceptual and practical advantages. To
begin, the method is
conceptually straightforward, relying only on the moment
structure of the
evolution equations and the smoothness of the quark
distributions. Since
these properties hold for all parton distributions the
method is applicable,
essentially without change, to both singlet and
nonsinglet distributions in both leading and next to
leading order in
$\alpha_s$. In addition, one need not commit to a
particular form for the
parton distributions when fitting data, but can instead
treat the moments
of the distribution, which are related to the physically
relevant twist two
matrix elements, as fit parameters.
Practically, the spline methods used here
are more stable than schemes involving orthogonal
polynomials,
which are prone to global oscillations when
singularities are encountered.
Since the elements of the scheme we have described,
numerical integration,
matrix inversion, and spline interpolation, are
extensively used in other
applications, the method is easily implemented. If
greater accuracy is desired,
the method can be altered to include any non-integral,
finite moment of the
parton distributions by simply generating the
appropriate anomalous dimensions
and coefficients. Since the matrix $A_{ni}$ need only be
inverted once to fit
data over a large range in $Q^2$,
the current method is faster than brute force
integration of
the Altarelli-Parisi equations.

\vfill\eject

\centerline{\bf {Acknowledgements}}
	This work was supported in part by the U.S.
Department of
Energy, Division of High Energy and Nuclear Physics,
ER-23.

\vfill\eject
\centerline{ Figure Captions}
\begin{itemize}
\item{}Figure 1 Reconstruction of the curve $\sqrt{x}
(1-x)^3$ using 3,5,7 and 10 splines.
\item{}Figure 2 Leading order evolution of the valence
$d$ quark distribution
of Morfin and Tung(MT)\cite{4}.
\end{itemize}
\end{document}